\documentclass[11pt,a4paper]{article}

\usepackage[T1]{fontenc}
\usepackage[utf8]{inputenc}
\usepackage{lmodern}
\usepackage[margin=2.6cm]{geometry}
\usepackage{amsmath,amssymb,amsthm}
\usepackage{graphicx}
\usepackage{booktabs}
\usepackage{tikz}
\usetikzlibrary{arrows.meta,positioning,calc,shapes.geometric}
\usepackage[square,comma,authoryear]{natbib}
\usepackage[hidelinks]{hyperref}
\usepackage{authblk}
\usepackage{caption}

\tikzset{
  ob/.style={draw, rectangle, align=flush left, inner sep=4pt, font=\small,
             text width=#1},
  ob/.default=3.1cm,
  asp/.style={-{Stealth[length=5pt]}, thin},
  isa/.style={{Hooks[right,length=3pt]}-{Stealth[length=5pt]}, thin},
  span/.style={-{Stealth[length=5pt]}, thin, dashed},
  lab/.style={font=\footnotesize, inner sep=2pt},
  gp/.style={draw, ellipse, font=\small, inner sep=2pt, fill=blue!8},
  term/.style={draw, rectangle, rounded corners=2pt, align=flush center,
               font=\small, fill=blue!8, text width=#1},
  term/.default=3.6cm,
}
\newcommand{\ck}{\checkmark}
\newcommand{\lt}[1]{\ifmmode\ulcorner\text{#1}\urcorner\else$\ulcorner$#1$\urcorner$\fi}

\newtheorem*{remark*}{Remark}

\title{Gene ologs: a categorical framework for Gene Ontology}
\author{Yanying Wu\thanks{\texttt{yanying.wu@cncb.ox.ac.uk}}}
\affil{Centre for Neural Circuits and Behaviour, University of Oxford, UK}
\date{21 November 2026}

\begin{document}
\maketitle

\begin{abstract}
The Gene Ontology (GO) is the most widely used resource for the functional
annotation of genes and gene products. It unifies biological knowledge across
species through a dynamic, controlled vocabulary, and it is now primarily
represented in the Web Ontology Language (OWL), a logic-based language that
supports automated reasoning. Ologs (ontology logs), introduced by Spivak and
Kent, offer a different language for knowledge representation, grounded in
category theory: concepts are \emph{types}, functional relationships are
\emph{aspects}, and declared equalities between paths are \emph{facts}. Ologs
come with features that are not part of the standard use of GO: path
equations, universal constructions (products, coproducts, pullbacks and
pushouts), a database reading in which instance data are set-valued functors,
and functors with which ontologies can be compared and merged. In this work we
take an initial step towards \emph{Gene ologs}, i.e.\ ologs for GO. We fix
conventions for reading GO terms as olog types, examine which GO relations can
be represented as aspects and which must instead be represented as spans,
illustrate each olog construction with GO examples, show that standard GO
annotations and GO Causal Activity Models (GO-CAM) can be represented as olog
instance data subject to checkable constraints, and describe how functors and
pushouts can be used to compare and merge Gene ologs. We also discuss what
ologs do not provide relative to OWL. We regard Gene ologs as complementary
to, rather than a replacement for, the OWL representation of GO.
\end{abstract}

\section{Introduction}

Despite the remarkable diversity of life on earth, genome sequencing has
revealed that many genes are conserved across species. Knowledge of gene
function gained in one species can therefore often be transferred to another.
Free-text descriptions of gene function, however, are ambiguous and are poorly
suited to such transfer. The Gene Ontology (GO) was created to fill this gap.
GO provides a dynamic, controlled vocabulary for annotating gene functions, so
that our understanding of different biological systems can be shared,
compared and computed with. The GO Consortium was founded in 1998, and GO has
since become the central resource for the unification of biology
\citep{Ashburner2000,GOC2019,GOC2026}.

GO was originally distributed in the Open Biomedical Ontologies (OBO) format,
and in recent years it has moved towards the Web Ontology Language (OWL)
\citep{Hastings2017}. OWL was standardised by the World Wide Web Consortium
(W3C) as part of the Semantic Web (OWL in 2004 and OWL~2 in 2009), and it is
supported by a large community and extensive tooling
\citep{Hastings2017,Hitzler2010}. OWL is based on description logics; its
axioms constrain classes and properties and allow consistency checking and
automated inference. Two limitations are, however, relevant here. First, the
full language (OWL~2 DL) has a high worst-case reasoning complexity, so large
biomedical ontologies such as GO rely on restricted profiles, notably OWL~2
EL, which achieve tractable reasoning partly by omitting constructs such as
class union (disjunction) \citep{Motik2012,Hitzler2010}. Second, OWL is a
two-valued logic and does not natively represent graded or uncertain knowledge
\citep{Hastings2017}. We note at the outset that ologs, as used here, are also
two-valued; our motivation for exploring them lies elsewhere.

Ologs were introduced by Spivak and colleagues \citep{Spivak2011} and
developed systematically by \citet{SpivakKent2012} as a category-theoretic
framework for knowledge representation. An olog is a category whose objects
and morphisms carry natural-language labels. Its mathematical foundation gives
ologs a precise semantics while keeping them readable. Three features
distinguish ologs from graph-based ontologies: (i) two paths between the same
types can be declared equal, which turns a graph into a category; (ii) limits
and colimits (products, pullbacks, coproducts, pushouts) construct new types
from old ones with a precise meaning; and (iii) functors relate different
ologs in a structure-preserving way, and set-valued functors give ologs a
database semantics \citep{SpivakKent2012,Spivak2012}. \citet{SpivakKent2012}
explicitly left a comparison between ologs and OWL to future work, and we do
not claim that ologs are more expressive than OWL. Rather, we ask what an
olog-based view of GO looks like and what it makes explicit.

The contributions of this paper are as follows. In Section~\ref{sec:go} we
review GO, its OWL representation, standard annotations and GO-CAM models. In
Section~\ref{sec:ologs} we review the basic notions of ologs, following
\citet{SpivakKent2012}. In Section~\ref{sec:geneologs} we draft Gene ologs:
we state how GO terms are read as types, check which GO relations qualify as
aspects, illustrate each olog construction with GO examples, model standard
annotations and GO-CAM models as olog instance data, and describe how Gene
ologs can be compared and merged. Section~\ref{sec:discussion} discusses the
relation to OWL and the limitations of the approach, and
Section~\ref{sec:conclusion} concludes.

\section{Gene Ontology and its OWL representation}\label{sec:go}

\subsection{GO and OWL basics}

GO consists of three disjoint sub-ontologies, or \emph{namespaces} (also
called GO aspects): biological process (BP), molecular function (MF) and
cellular component (CC).\footnote{To avoid a clash with the olog notion of an
\emph{aspect} (Section~\ref{sec:ologs}), we use the word \emph{namespace} for
the three branches of GO throughout this paper.}

A \emph{biological process} is a biological programme carried out by one or
more ordered assemblies of molecular functions. For example, ``cellular ion
homeostasis'' (GO:0006873) refers to any process involved in maintaining an
internal steady state of ions at the level of a cell. The \emph{Drosophila}
gene \textit{DAT}, which encodes a dopamine transporter, is annotated to this
biological process. A \emph{molecular function} is an activity performed at
the molecular level by a gene product; \textit{DAT} is annotated, for example,
to ``dopamine:sodium symporter activity'' (GO:0005330). A \emph{cellular
component} is a location, relative to cellular structures, at which a gene
product performs its function; one of the cellular-component annotations of
\textit{DAT} is ``integral component of plasma membrane''
(GO:0005887).\footnote{GO term labels change between releases, and terms may
be merged or made obsolete. Labels in the \textit{DAT} and \textit{Foxp1}
examples are those used when the annotations were retrieved for the original
version of this work (2019); GO identifiers should be used to resolve terms
against a given release.} Taken together, these three kinds of annotation give
a picture of what a gene product does: \textit{DAT} encodes a
dopamine:sodium symporter that is located in the plasma membrane and is
involved in cellular ion homeostasis.

GO is now primarily represented in OWL \citep{Hastings2017,Hitzler2010}. The
main components of an OWL ontology are classes, properties, individuals and
restrictions: classes denote concepts, properties denote relations, individuals
are instances of classes, and restrictions give logical meaning to class
definitions. OWL supports quantification (existential and universal),
cardinality restrictions (exact, minimum, maximum), the Boolean connectives
(intersection, union, complement), and disjointness and equivalence of
classes, although, as noted above, profiles such as OWL~2 EL support only part
of this language \citep{Motik2012}. These logical constructs enable automated
reasoning: the consistency of GO can be checked and implicit subclass
relationships can be inferred.

GO terms and the relations between them form a directed acyclic graph.
Figure~\ref{fig:goexample} shows a simplified fragment around the term
``negative regulation of DNA-templated transcription'' (GO:0045892). Each box
is a class (GO term), and each arrow is a relation between classes, with the
relation types given in the legend \citep{Binns2009,Carbon2009}. Note that
GO relations between classes have an ``all--some'' reading: for instance,
``$A$ negatively regulates $B$'' means that every instance of $A$ negatively
regulates some instance of $B$.

\begin{figure}[ht]
\centering
\begin{tikzpicture}[node distance=1.2cm]
  \node[term=5.0cm] (bp) at (0,0) {biological process\\[-1pt]\footnotesize GO:0008150};
  \node[term=5.0cm] (cp) at (0,-1.8) {cellular process\\[-1pt]\footnotesize GO:0009987};
  \node[term=5.0cm] (nrcp) at (0,-3.6) {negative regulation of cellular process\\[-1pt]\footnotesize GO:0048523};
  \node[term=5.0cm] (nrt) at (0,-5.7) {negative regulation of DNA-templated transcription\\[-1pt]\footnotesize GO:0045892};
  \draw[-{Stealth[length=5pt]}, thick] (cp) -- (bp);
  \draw[-{Stealth[length=5pt]}, thick, red!75!black] (nrcp) -- (cp);
  \draw[-{Stealth[length=5pt]}, thick, densely dotted] (nrt) -- (nrcp);
  \begin{scope}[shift={(3.4,-0.6)}, font=\footnotesize]
    \draw[-{Stealth[length=5pt]}, thick] (0,0) -- (1.2,0) node[right] {is a};
    \draw[-{Stealth[length=5pt]}, thick, densely dotted] (0,-0.5) -- (1.2,-0.5) node[right, align=left] {is a (path; intermediate terms omitted)};
    \draw[-{Stealth[length=5pt]}, thick, red!75!black] (0,-1.0) -- (1.2,-1.0) node[right] {negatively regulates};
  \end{scope}
\end{tikzpicture}
\caption{A simplified fragment of GO (based on AmiGO/QuickGO). The dotted
arrow stands for a chain of ``is a'' relations whose intermediate terms are
omitted; other relations of these terms are also omitted.}
\label{fig:goexample}
\end{figure}

A particular gene product, such as the transcription factor Foxp1, can be
annotated with ``negative regulation of DNA-templated transcription'' on the
basis of experimental evidence \citep{Li2004}. As of the GO release of 19 June
2026, the GO knowledgebase contained 38,263 GO terms and 14,736,040
annotations for 1,750,636 gene products.\footnote{Release statistics as reported for GO release
2026-06-19 (\url{https://release.geneontology.org/2026-06-19/}).}

\subsection{Standard GO annotation and GO-CAM models}\label{sec:gocam}

In a standard GO annotation, a gene product is associated with a single GO
term, via a relation that depends on the namespace of the term (e.g.\
``enables'' for molecular functions; ``involved in'' or ``acts upstream of or
within'' for biological processes; ``located in'' or ``part of'' for cellular
components), together with supporting evidence and a reference. Many GO terms
can be assigned to one gene product, and one GO term is typically associated
with many gene products. In both directions the associated items are
biologically related: several gene products annotated with the same term may
act together in a pathway, and the molecular function of a gene product is
related to the process it contributes to and the location in which it acts.
Standard annotations do not record these connections. GO Causal Activity
Models (GO-CAM) were introduced to do so: a GO-CAM model links molecular
activities, the gene products that enable them, the processes they are part
of or causally upstream of, and the locations where they occur, into a single
annotation network \citep{Thomas2019,GOC2019}.

\begin{figure}[ht]
\centering
\includegraphics[width=0.62\textwidth]{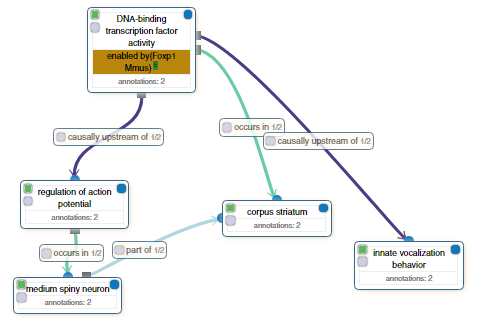}
\caption{The GO-CAM model for Foxp1, as rendered by Noctua
(geneontology.org).}
\label{fig:foxp1-noctua}
\end{figure}

As an example, Figure~\ref{fig:foxp1-std} shows some of the individual GO
annotations of mouse Foxp1 (not a complete list), and
Figures~\ref{fig:foxp1-cam} and \ref{fig:foxp1-noctua} show the corresponding
GO-CAM model, as a schematic and as rendered by the GO-CAM editor Noctua
(\url{http://noctua.geneontology.org}). The model is based on
\citet{Araujo2015}. Note that the GO-CAM model also refers to an anatomical
structure (corpus striatum) and a cell type (medium spiny neuron). These are
not GO terms; they come from the Uberon anatomy ontology and the Cell Ontology,
respectively, and in standard annotations such information can only appear as
an annotation extension.

\begin{figure}[ht]
\centering
\begin{tikzpicture}
  \node[gp] (g) at (0,0) {Foxp1};
  \node[term=5.4cm, anchor=west] (mf) at (5.6,1.5) {DNA-binding transcription factor activity (MF)};
  \node[term=5.4cm, anchor=west] (bp1) at (5.6,0) {regulation of action potential (BP)};
  \node[term=5.4cm, anchor=west] (bp2) at (5.6,-1.5) {innate vocalization behavior (BP)};
  \draw[asp] (g.north east) .. controls (1.5,1.5) .. node[lab, above, pos=0.78] {enables} (mf.west);
  \draw[asp] (g.east) -- node[lab, above, pos=0.55] {acts upstream of or within} (bp1.west);
  \draw[asp] (g.south east) .. controls (1.5,-1.5) .. node[lab, below, pos=0.78] {acts upstream of or within} (bp2.west);
\end{tikzpicture}
\caption{Some standard GO annotations of Foxp1. Each annotation relates a
gene product to a single GO term; the relation depends on the namespace of
the term. The connections between these annotations are not recorded.}
\label{fig:foxp1-std}
\end{figure}

\begin{figure}[ht]
\centering
\begin{tikzpicture}
  \node[gp] (g) at (0,0) {Foxp1};
  \node[term=4.0cm] (mf) at (5.0,0) {DNA-binding transcription factor activity};
  \node[term=3.2cm] (cs) at (11.0,0) {corpus striatum};
  \node[term=3.2cm] (rap) at (5.0,-2.6) {regulation of action potential};
  \node[term=3.2cm] (msn) at (11.0,-2.6) {medium spiny neuron};
  \node[term=3.2cm] (ivb) at (0.6,-2.6) {innate vocalization behavior};
  \draw[asp, green!45!black] (mf) -- node[lab, above] {enabled by} (g);
  \draw[asp] (mf) -- node[lab, above] {occurs in} (cs);
  \draw[asp] (mf) -- node[lab, right] {causally upstream of} (rap);
  \draw[asp] (mf.south west) -- node[lab, left, align=right, pos=0.45] {causally\\upstream of} (ivb.north east);
  \draw[asp] (rap) -- node[lab, above] {occurs in} (msn);
  \draw[asp] (msn) -- node[lab, right] {part of} (cs);
\end{tikzpicture}
\caption{Schematic of the GO-CAM model for Foxp1 (after
Figure~\ref{fig:foxp1-noctua}). The activity is enabled by Foxp1; it occurs in the corpus striatum and is
causally upstream of two biological processes. Corpus striatum (Uberon) and
medium spiny neuron (Cell Ontology) are not GO terms.}
\label{fig:foxp1-cam}
\end{figure}

From Figure~\ref{fig:foxp1-std} we learn several isolated facts about Foxp1;
from Figures~\ref{fig:foxp1-cam} and \ref{fig:foxp1-noctua} we learn how they
fit together. In Section~\ref{sec:instances} we show that both kinds of
information can be represented as instance data of Gene ologs, and that the
olog representation makes explicit constraints that the data must satisfy.

\section{Ologs for knowledge representation}\label{sec:ologs}

The olog (ontology log) is a language for ontologies that differs markedly from
OWL. It was introduced by Spivak and colleagues in 2011
\citep{Spivak2011,SpivakKent2012}. The central idea is to use category theory as
a framework for knowledge representation. Category theory is a branch of
mathematics introduced by \citet{EilenbergMacLane1945} to relate algebra and
topology (specifically, algebraic topology). Since then it has proved
effective at unifying seemingly separate disciplines and at revealing the
connections and similarities between them, and it has found applications in
physics, computer science and linguistics \citep{Spivak2014,Awodey2010,Sica2006}.
It has also been proposed as a general language for qualitative modelling in
science \citep{Spivak2011,Spivak2014}. At the core of category theory is
composition: how simple building blocks combine into complex structures in a
controlled way. This matches a central task of an ontology language, which is
to combine pieces of knowledge coherently.

An olog is a category presented in natural language. Its objects are called
\emph{types} and its morphisms \emph{aspects}. A type is drawn as a box
containing a singular indefinite noun phrase such as \lt{a gene}; it stands
for the set of things of that type. An aspect of a type $X$ is a way of
viewing or measuring each instance of $X$, and it is drawn as a labelled arrow
$X\to Y$, where $Y$ is the set of possible results. For example, an olog
recording that a gene is a segment of DNA is
\begin{equation}\label{olog:gene}
\begin{tikzpicture}[baseline=(current bounding box.center)]
  \node[ob=1.6cm] (g) at (0,0) {a gene};
  \node[ob=3.0cm] (d) at (5,0) {a segment of DNA};
  \draw[isa] (g) -- node[lab, above] {is} (d);
\end{tikzpicture}
\end{equation}
Following \citet{SpivakKent2012}, an olog should be read as the world-view of
its author; olog~(\ref{olog:gene}), for instance, ignores genes with RNA
genomes.

\paragraph{Aspects are functional.} For aspects to be composable, ologs require
them to be \emph{functional relationships}: if the types $A$ and $B$ are
regarded as sets, an aspect $f\colon A\to B$ must assign to each element of
$A$ exactly one element of $B$ \citep{SpivakKent2012}. $A$ is called the
\emph{domain} of $f$ and $B$ its \emph{codomain} (\citealt{SpivakKent2012}
also call $B$ the ``set of result values'' of $f$). Given another aspect
$g\colon B\to C$, the two compose to $g\circ f\colon A\to C$. Arrows that are
not functional are invalid as aspects. For example,
$\lt{a person}\xrightarrow{\text{has}}\lt{a child}$ is not an aspect, since a
person may have no children or several. Such relations can often be made
functional by reversing them
($\lt{a person}\xrightarrow{\text{has as mother}}\lt{a woman}$) or by
replacing them with a \emph{span}: a new type whose instances are the related
tuples, together with one aspect to each participating type
\citep[Section~2.2.3]{SpivakKent2012}. Aspects labelled ``is'' or ``is a'' are
moreover \emph{injective} (different instances remain different), and we draw
them with hooked arrows $\hookrightarrow$.

\paragraph{Facts.} An olog may declare that two paths with the same source and
target are equivalent; such a declaration is a \emph{fact}, and the diagram is
then said to \emph{commute}. Being able to declare path equivalences is
precisely what distinguishes a category from a graph \citep{SpivakKent2012}.
Commutativity is indicated by a check mark (\ck) in the region bounded by the
two paths. Consider
\begin{equation}\label{olog:dogma}
\begin{tikzpicture}[baseline=(current bounding box.center)]
  \node[ob=3.3cm] (t) at (0,0) {a protein-coding transcript};
  \node[ob=3.0cm] (p) at (8.4,0) {a protein isoform};
  \node[ob=1.4cm] (g) at (4.2,-2.5) {a gene};
  \draw[asp] (t) -- node[lab, above] {has as translation product} (p);
  \draw[asp] (t.south) -- node[lab, below left, pos=0.45] {is transcribed from} (g.west);
  \draw[asp] (p.south) -- node[lab, below right, pos=0.45] {is encoded by} (g.east);
  \node at (4.2,-0.9) {\ck};
\end{tikzpicture}
\end{equation}
All three arrows are functional, provided that the author regards each
protein-coding transcript as having a single translation product and each
transcript and each protein isoform as deriving from a single gene (which
ignores, for example, polycistronic and trans-spliced transcripts). The check
mark declares the fact that \emph{the gene from which a transcript is
transcribed is the gene that encodes the protein isoform into which the
transcript is translated}. Note the direction of the arrows. The version in
which the arrows start at \lt{a gene} (``a gene can be transcribed into an
mRNA'', ``a gene encodes a protein'') is not a valid olog, because a gene may
have several transcripts and protein isoforms, and non-coding genes have none.

\paragraph{Instances.} Types, aspects and facts all have instances
\citep[Section~3]{SpivakKent2012}. An olog can be read as a database schema:
each type becomes a table whose rows are its documented instances, each
aspect becomes a column holding, for every row, an instance of the codomain,
and each fact becomes an integrity constraint. Formally, instance data on an
olog $\mathcal{C}$ is a functor $I\colon\mathcal{C}\to\mathbf{Set}$, which
assigns a set $I(X)$ to each type $X$ and a function $I(f)\colon I(X)\to I(Y)$
to each aspect $f\colon X\to Y$, such that every declared fact holds as an
equality of functions \citep{SpivakKent2012,Spivak2012}.

\paragraph{Universal constructions.} What makes ologs considerably more
expressive than labelled graphs is that types can be \emph{constructed} from
other types by limits (such as products and pullbacks) and colimits (such as
coproducts and pushouts) \citep[Sections~5--6]{SpivakKent2012}. Each such
construction is characterised by a universal property, so its meaning is
fixed by the rest of the olog rather than by its label alone. We describe
these constructions, with GO examples, in Section~\ref{sec:constructions}.

\section{Gene ologs: an initial draft}\label{sec:geneologs}

In this section we draft an initial version of Gene ologs. We adapt
definitions from the olog framework \citep{SpivakKent2012} to GO and
illustrate each with examples.

\subsection{Basic components}\label{sec:basic}

\paragraph{Reading GO terms as types.} A type is a concept, drawn as a box
containing a singular indefinite noun phrase. In a Gene olog most types come
from GO terms: we prefix the term label with ``a'' or ``an'' (rephrasing the
label slightly if needed). As in GO's own OWL semantics, the set associated
with a GO-term type consists of the \emph{instances of that class}, not of
gene products:
\begin{itemize}
  \item an instance of a BP type, e.g.\ \lt{a biosynthetic process}, is a
  particular occurrence of that process;
  \item an instance of an MF type, e.g.\ \lt{a protein kinase activity}, is a
  particular occurrence of that activity, which is enabled by some gene
  product;
  \item an instance of a CC type, e.g.\ \lt{a ribosome}, is a particular
  structure of that kind.
\end{itemize}
Gene products form a separate type, \lt{a gene product}, which is not a GO
term; Section~\ref{sec:instances} describes how it is connected to GO-term
types. Keeping these readings apart matters: \lt{a GTPase activator activity}
(a kind of activity) and \lt{a GTPase activator} (a kind of protein) are
different types. Typical types are
\begin{equation}\label{olog:types}
\begin{tikzpicture}[baseline=(current bounding box.center)]
  \node[ob=3.3cm] at (0,0) {a biosynthetic process};
  \node[ob=3.3cm] at (4.4,0) {a protein kinase activity};
  \node[ob=2.2cm] at (8.2,0) {a ribosome};
  \node[ob=2.4cm] at (11.2,0) {a gene product};
\end{tikzpicture}
\end{equation}
corresponding to GO:0009058 (BP), GO:0004672 (MF), GO:0005840 (CC), and the
non-GO type of gene products.

\paragraph{Aspects.} An aspect is drawn as an arrow between two types with a
label that begins with a verb, so that reading source, label and target gives
an English sentence. Because aspects must be functional, a GO relation can
only be used as an aspect after checking that, in the author's world-view,
every instance of the source is related to \emph{exactly one} instance of the
target. GO's ``all--some'' reading guarantees \emph{at least} one; uniqueness
is an additional modelling commitment. Examples are
\begin{equation}\label{olog:aspects}
\begin{tikzpicture}[baseline=(current bounding box.center)]
  \node[ob=4.6cm] (a1) at (0,0) {a protein serine/threonine kinase activity};
  \node[ob=3.4cm] (b1) at (7.2,0) {a protein kinase activity};
  \draw[isa] (a1) -- node[lab, above] {is a} (b1);
  \node[ob=4.6cm] (a2) at (0,-1.3) {a Golgi apparatus};
  \node[ob=3.4cm] (b2) at (7.2,-1.3) {an endomembrane system};
  \draw[asp] (a2) -- node[lab, above] {is part of} (b2);
  \node[ob=4.6cm] (a3) at (0,-2.7) {a negative regulation of DNA-templated transcription};
  \node[ob=3.4cm] (b3) at (7.2,-2.7) {a DNA-templated transcription};
  \draw[asp] (a3) -- node[lab, above] {negatively regulates} (b3);
\end{tikzpicture}
\end{equation}
The ``is a'' relation of GO always yields an injective aspect, interpreted as
the inclusion of one set of instances into another. The second arrow is
functional because each Golgi apparatus is part of the endomembrane system of
exactly one cell. The third is functional if each occurrence of negative
regulation of transcription is regarded as acting on a single transcription
process. By contrast, an arrow such as
$\lt{a transcription factor}\xrightarrow{\text{regulates}}\lt{a gene}$ is
\emph{not} an aspect, because a transcription factor regulates many genes and
a gene is regulated by many transcription factors; such many-to-many relations
must be represented as spans (Section~\ref{sec:instances}).

\paragraph{Paths.} When the codomain of one aspect is the domain of another,
the two can be joined into a path, which denotes their composite:
\begin{equation}\label{olog:path}
\begin{tikzpicture}[baseline=(current bounding box.center)]
  \node[ob=3.3cm] (a) at (0,0) {a GTPase activator activity};
  \node[ob=3.3cm] (b) at (6,0) {an enzyme activator activity};
  \node[ob=3.3cm] (c) at (6,-1.6) {an enzyme regulator activity};
  \draw[isa] (a) -- node[lab, above] {is an} (b);
  \draw[isa] (b) -- node[lab, right] {is an} (c);
\end{tikzpicture}
\end{equation}

\paragraph{Facts.} A fact is a declaration that two paths with the same
source and target are equivalent, indicated by a check mark. A fact often
expresses that a GO term can be viewed in two ways. For example,
\begin{equation}\label{olog:gtpase}
\begin{tikzpicture}[baseline=(current bounding box.center)]
  \node[ob=3.3cm] (a) at (0,0) {a GTPase activator activity};
  \node[ob=3.3cm] (b) at (6,0) {an enzyme activator activity};
  \node[ob=3.3cm] (c) at (0,-1.8) {a GTPase regulator activity};
  \node[ob=3.3cm] (d) at (6,-1.8) {an enzyme regulator activity};
  \draw[isa] (a) -- node[lab, above] {is an} (b);
  \draw[isa] (a) -- node[lab, left] {is a} (c);
  \draw[isa] (b) -- node[lab, right] {is an} (d);
  \draw[isa] (c) -- node[lab, above] {is an} (d);
  \node at (3,-0.9) {\ck};
\end{tikzpicture}
\end{equation}
states that a GTPase activator activity is an enzyme regulator activity
whether it is viewed as a GTPase regulator activity or as an enzyme activator
activity (in GO, the bottom arrow is a path through ``nucleoside-triphosphatase
regulator activity'').

\begin{remark*}
Olog~(\ref{olog:gtpase}) commutes automatically: every ``is a'' aspect is an
inclusion of instance sets, and any two composites of inclusions with the same
source and target are the same inclusion. A diagram built solely from GO's
``is a'' hierarchy is therefore a preorder, in which every diagram commutes,
and the check mark records no new information. Facts carry real content only
when other aspects are involved, as in olog~(\ref{olog:dogma}) and in the
annotation constraint of olog~(\ref{olog:annotation}) below. Following
\citet[Rule~2.3.4]{SpivakKent2012}, we nevertheless declare such facts
explicitly.
\end{remark*}

\paragraph{Instances.} Since a type represents a set, an instance of a type is
an element of that set. As explained above, the instances of GO-term types are
occurrences of processes and activities, and particular cellular structures.
In particular, \emph{annotating a gene product with a GO term does not make the
gene product an instance of that term}: the gene product Foxp1 is not an
instance of \lt{a DNA-binding transcription factor activity}, but enables
instances of it. Annotations are therefore modelled as a relation between
gene products and GO terms (Section~\ref{sec:instances}). Formally, instance
data on a Gene olog $\mathcal{C}$ is a functor $I\colon\mathcal{C}\to\mathbf{Set}$
(Section~\ref{sec:ologs}).

\subsection{Universal constructions}\label{sec:constructions}

\paragraph{Products.} The product of two types (sets) $A$ and $B$, denoted
$A\times B$, is the set of pairs
\[
A\times B=\{(a,b)\mid a\in A \text{ and } b\in B\},
\]
together with the two projection aspects $A\times B\to A$ and $A\times B\to B$
sending $(a,b)$ to $a$ and to $b$, respectively. The product is characterised
by a universal property: for any type $X$ with aspects $f\colon X\to A$ and
$g\colon X\to B$ there is a unique aspect $\langle f,g\rangle\colon X\to A\times
B$ whose composites with the projections are $f$ and $g$. For example, let
\lt{an annotation} be the type of GO annotations of
Section~\ref{sec:instances}, each of which has a gene product and a GO term.
The universal property yields the dashed aspect below, and both triangles
commute:
\begin{equation}\label{olog:product}
\begin{tikzpicture}[baseline=(current bounding box.center)]
  \node[ob=2.6cm] (x) at (0,0) {an annotation};
  \node[ob=4.6cm] (p) at (0,-2.0) {a pair $(g,t)$ where $g$ is a gene product and $t$ is a GO term};
  \node[ob=2.4cm] (a) at (-5.2,-2.0) {a gene product};
  \node[ob=2.0cm] (b) at (5.0,-2.0) {a GO term};
  \draw[asp] (x) -- node[lab, above, sloped] {is of} (a);
  \draw[asp] (x) -- node[lab, above, sloped] {is to} (b);
  \draw[asp, dashed] (x) -- node[lab, right] {yields} (p);
  \draw[asp] (p) -- node[lab, below] {$g$} (a);
  \draw[asp] (p) -- node[lab, below] {$t$} (b);
  \node at (-2.3,-1.25) {\ck};
  \node at (2.3,-1.25) {\ck};
\end{tikzpicture}
\end{equation}
The dashed aspect need not be injective: a gene product may be annotated to
the same term more than once, with different evidence. This is why
annotations are better represented by a span than by a subset of the product.

\paragraph{Coproducts.} The coproduct of two sets $A$ and $B$, denoted
$A\sqcup B$, is their disjoint union
\[
A\sqcup B=\{(a,\text{``}A\text{''})\mid a\in A\}\cup\{(b,\text{``}B\text{''})\mid b\in B\},
\]
together with the inclusion aspects $A\to A\sqcup B$ and $B\to A\sqcup B$
sending $a$ to $(a,\text{``}A\text{''})$ and $b$ to $(b,\text{``}B\text{''})$.
Dually to the product, for any type $X$ with aspects $A\to X$ and $B\to X$
there is a unique aspect $A\sqcup B\to X$ that restricts to them. Because the
three GO namespaces are disjoint, the set of GO terms is their coproduct, and
the aspect assigning each term its namespace is induced by the universal
property from the three constant aspects:
\begin{equation}\label{olog:coproduct}
\begin{tikzpicture}[baseline=(current bounding box.center)]
  \node[ob=2.4cm] (mf) at (-4.2,0) {an MF term};
  \node[ob=2.4cm] (bp) at (0,0) {a BP term};
  \node[ob=2.4cm] (cc) at (4.2,0) {a CC term};
  \node[ob=5.4cm] (t) at (0,-2.0) {a GO term (an MF term, a BP term or a CC term)};
  \node[ob=4.4cm] (ns) at (0,-3.9) {a GO namespace (one of MF, BP, CC)};
  \draw[isa] (mf.south) -- node[lab, above, sloped] {is a} (t);
  \draw[isa] (bp) -- node[lab, right] {is a} (t);
  \draw[isa] (cc.south) -- node[lab, above, sloped] {is a} (t);
  \draw[asp] (t) -- node[lab, right] {belongs to} (ns);
\end{tikzpicture}
\end{equation}
Here $\lt{a GO term}=\lt{an MF term}\sqcup\lt{a BP term}\sqcup\lt{a CC term}$,
and ``belongs to'' sends every MF term to MF, every BP term to BP and every CC
term to CC.

\paragraph{Pullbacks.} Let $f\colon B\to D$ and $g\colon C\to D$ be aspects.
Their pullback, denoted $B\times_D C$, is the type
\[
B\times_D C=\{(b,c)\mid b\in B,\ c\in C \text{ and } f(b)=g(c)\},
\]
together with the projection aspects to $B$ and $C$; in olog~(\ref{olog:pullback})
the right-hand square is the pullback of the left-hand diagram:
\begin{equation}\label{olog:pullback}
\begin{tikzpicture}[baseline=(current bounding box.center)]
  \node[ob=1.6cm] (C) at (1.8,0) {type $C$};
  \node[ob=1.6cm] (B) at (-1.2,-1.8) {type $B$};
  \node[ob=1.6cm] (D) at (1.8,-1.8) {type $D$};
  \draw[asp] (B) -- node[lab, above] {$f$} (D);
  \draw[asp] (C) -- node[lab, right] {$g$} (D);
  \begin{scope}[xshift=6.2cm]
  \node[ob=2.1cm] (P) at (0,0) {type $B\times_D C$};
  \node[ob=1.6cm] (C2) at (3.4,0) {type $C$};
  \node[ob=1.6cm] (B2) at (0,-1.8) {type $B$};
  \node[ob=1.6cm] (D2) at (3.4,-1.8) {type $D$};
  \draw[asp] (P) -- node[lab, above] {$p_C$} (C2);
  \draw[asp] (P) -- node[lab, left] {$p_B$} (B2);
  \draw[asp] (B2) -- node[lab, above] {$f$} (D2);
  \draw[asp] (C2) -- node[lab, right] {$g$} (D2);
  \node at (1.7,-0.9) {\ck};
  \end{scope}
\end{tikzpicture}
\end{equation}
Commutativity of the square ($f\circ p_B=g\circ p_C$) is necessary but not
sufficient. What makes the square a pullback is its universal property: for
every type $X$ with aspects $x_B\colon X\to B$ and $x_C\colon X\to C$ such that
$f\circ x_B=g\circ x_C$, there is a unique aspect $X\to B\times_D C$ through
which $x_B$ and $x_C$ factor. (Without this condition, for example, the empty
type would also make the square commute.)

When $f$ and $g$ are both ``is a'' aspects, the pullback is the intersection
of $B$ and $C$, and so pullbacks express ``and''. In GO, ``carbohydrate
transmembrane transport'' (GO:0034219) has exactly two parents, ``carbohydrate
transport'' (GO:0008643) and ``transmembrane transport'' (GO:0055085):
\begin{equation}\label{olog:pb-go}
\begin{tikzpicture}[baseline=(current bounding box.center)]
  \node[ob=4.0cm] (P) at (0,0) {a carbohydrate transmembrane transport};
  \node[ob=3.2cm] (C) at (6.2,0) {a transmembrane transport};
  \node[ob=3.2cm] (B) at (0,-2.0) {a carbohydrate transport};
  \node[ob=1.9cm] (D) at (6.2,-2.0) {a transport};
  \draw[isa] (P) -- node[lab, above] {is a} (C);
  \draw[isa] (P) -- node[lab, left] {is a} (B);
  \draw[isa] (B) -- node[lab, above] {is a} (D);
  \draw[isa] (C) -- node[lab, right] {is a} (D);
  \node at (3.1,-1.0) {\ck};
\end{tikzpicture}
\end{equation}
Declaring this square to be a pullback asserts more than the two ``is a''
relations: it asserts that \emph{every} transport that is both a carbohydrate
transport and a transmembrane transport is a carbohydrate transmembrane
transport. This is the olog counterpart of an OWL equivalence axiom defining
a class as an intersection. Note that the instances of all four types are
transport \emph{processes}, not gene products. Pullbacks can express many
other relationships as well \citep[Section~5.2]{SpivakKent2012}.

\paragraph{Pushouts.} Let $f\colon A\to B$ and $g\colon A\to C$ be aspects.
Their pushout, denoted $B\sqcup_A C$, is the quotient of the coproduct
$B\sqcup C$ by the equivalence relation generated by $f(a)\sim g(a)$ for every
$a\in A$ \citep[Definition~6.2.2]{SpivakKent2012}, together with the two
induced aspects into it:
\begin{equation}\label{olog:pushout}
\begin{tikzpicture}[baseline=(current bounding box.center)]
  \node[ob=1.6cm] (A) at (0,0) {type $A$};
  \node[ob=1.6cm] (C) at (3.0,0) {type $C$};
  \node[ob=1.6cm] (B) at (0,-1.8) {type $B$};
  \draw[asp] (A) -- node[lab, above] {$g$} (C);
  \draw[asp] (A) -- node[lab, left] {$f$} (B);
  \begin{scope}[xshift=6.0cm]
  \node[ob=1.6cm] (A2) at (0,0) {type $A$};
  \node[ob=1.6cm] (C2) at (3.4,0) {type $C$};
  \node[ob=1.6cm] (B2) at (0,-1.8) {type $B$};
  \node[ob=2.1cm] (Q) at (3.4,-1.8) {type $B\sqcup_A C$};
  \draw[asp] (A2) -- node[lab, above] {$g$} (C2);
  \draw[asp] (A2) -- node[lab, left] {$f$} (B2);
  \draw[asp] (B2) -- node[lab, above] {$i_B$} (Q);
  \draw[asp] (C2) -- node[lab, right] {$i_C$} (Q);
  \node at (1.7,-0.9) {\ck};
  \end{scope}
\end{tikzpicture}
\end{equation}
Again, commutativity is necessary but not sufficient; the pushout is
characterised by the dual universal property: for every type $X$ with aspects
$B\to X$ and $C\to X$ that agree on $A$, there is a unique aspect
$B\sqcup_A C\to X$ through which they factor.

Pushouts can express ``or'' without double counting. If $f$ and $g$ are ``is
a'' aspects and $A$ is \emph{the whole} intersection of $B$ and $C$ (that is,
$A=B\times_D C$ for some common supertype $D$), then $B\sqcup_A C$ is the union
of $B$ and $C$. Continuing the example of olog~(\ref{olog:pb-go}),
\begin{equation}\label{olog:po-go}
\begin{tikzpicture}[baseline=(current bounding box.center)]
  \node[ob=4.0cm] (A) at (0,0) {a carbohydrate transmembrane transport};
  \node[ob=3.2cm] (C) at (6.2,0) {a transmembrane transport};
  \node[ob=3.2cm] (B) at (0,-2.0) {a carbohydrate transport};
  \node[ob=4.3cm] (Q) at (6.6,-2.0) {a carbohydrate transport or a transmembrane transport};
  \draw[isa] (A) -- node[lab, above] {is a} (C);
  \draw[isa] (A) -- node[lab, left] {is a} (B);
  \draw[isa] (B) -- node[lab, above] {is a} (Q);
  \draw[isa] (C) -- node[lab, right] {is a} (Q);
  \node at (3.1,-1.0) {\ck};
\end{tikzpicture}
\end{equation}
Each carbohydrate transmembrane transport is counted once in the pushout, even
though it is both a carbohydrate transport and a transmembrane transport; in
the coproduct $B\sqcup C$ it would be counted twice. Two conditions matter
here. The arrows must be ``is a'' (inclusion) aspects: if they are, for
example, ``is part of'' aspects, the pushout glues wholes together along their
shared parts and does not form a union. And the apex must be the entire
overlap: if it contained only some of the common instances, the remaining
ones would still be counted twice. The pushout type in
olog~(\ref{olog:po-go}) has no counterpart in GO. Class unions of this kind
are expressible in OWL~2 DL but not in the OWL~2 EL profile
\citep{Motik2012}.

\subsection{Annotations and GO-CAM models as instance data}\label{sec:instances}

We now use the constructions above to give the annotations and GO-CAM models
of Section~\ref{sec:gocam} a place in Gene ologs.

\paragraph{Standard annotations as a span.} An annotation relates a gene
product to a GO term through a namespace-specific relation and is supported by
evidence. Since this relationship is many-to-many, it is modelled as a span:
\begin{equation}\label{olog:annotation}
\begin{tikzpicture}[baseline=(current bounding box.center)]
  \node[ob=5.6cm] (ann) at (0,0) {an annotation $(g,r,t)$ where gene product $g$ stands in relation $r$ to GO term $t$};
  \node[ob=2.4cm] (gp) at (-5.6,-2.6) {a gene product};
  \node[ob=3.9cm] (rel) at (-1.7,-2.6) {a gene-product-to-term relation};
  \node[ob=2.0cm] (term) at (2.6,-2.6) {a GO term};
  \node[ob=2.4cm] (ev) at (5.8,-2.6) {an evidence code};
  \node[ob=4.4cm] (ns) at (0.4,-4.9) {a GO namespace (one of MF, BP, CC)};
  \draw[asp] (ann.south west) -- node[lab, above left] {$g$} (gp.north);
  \draw[asp] (ann.south -| rel.north) -- node[lab, left] {$r$} (rel.north);
  \draw[asp] (ann.south -| term.north) -- node[lab, right] {$t$} (term.north);
  \draw[asp] (ann.south east) -- node[lab, above right] {is supported by} (ev.north);
  \draw[asp] (rel.south) -- node[lab, below left] {applies to terms of} (ns.north west);
  \draw[asp] (term.south) -- node[lab, below right] {belongs to} (ns.north east);
  \node at (0.4,-3.6) {\ck};
\end{tikzpicture}
\end{equation}
The check mark declares a fact with real content: \emph{the namespace of the
annotated term is the namespace to which the relation applies}. Thus
``enables'' may only be used with MF terms, ``involved in'' and ``acts
upstream of or within'' only with BP terms, and ``located in'' and ``part of''
only with CC terms. Any instance $I$ of olog~(\ref{olog:annotation}) must
satisfy
\[
I(\text{belongs to})\circ I(t)=I(\text{applies to terms of})\circ I(r),
\]
which is an integrity constraint that can be checked automatically on an
annotation file. Table~\ref{tab:annotation} shows part of an instance for the
annotations of Figure~\ref{fig:foxp1-std}; the last two columns evaluate the
two paths of the fact, and they agree on every row.

\begin{table}[ht]
\centering\small
\begin{tabular}{@{}l l l p{5.6cm} c c@{}}
\toprule
 & & & & \multicolumn{2}{c}{namespace} \\
\cmidrule(l){5-6}
annotation & $g$ & $r$ & $t$ & via $t$ & via $r$ \\
\midrule
$\alpha_1$ & Foxp1 & enables & DNA-binding transcription factor activity & MF & MF \\
$\alpha_2$ & Foxp1 & acts upstream of or within & regulation of action potential & BP & BP \\
$\alpha_3$ & Foxp1 & acts upstream of or within & innate vocalization behavior & BP & BP \\
\bottomrule
\end{tabular}
\caption{Part of an instance of olog~(\ref{olog:annotation}) for the
annotations of Figure~\ref{fig:foxp1-std} (evidence column omitted). The last
two columns evaluate the two paths of the declared fact.}
\label{tab:annotation}
\end{table}

\paragraph{GO-CAM models as instances.} In a GO-CAM model the nodes are
individuals: a particular activity, a particular process, a particular
anatomical location \citep{Thomas2019}. In the olog view, a GO-CAM model is
therefore not a new ontology but \emph{instance data} on a fixed schema olog,
part of which is
\begin{equation}\label{olog:gocam}
\begin{tikzpicture}[baseline=(current bounding box.center)]
  \node[ob=2.4cm] (gp) at (-5.2,0) {a gene product};
  \node[ob=2.4cm] (act) at (0,0) {an activity};
  \node[ob=2.0cm] (mf) at (5.0,0) {an MF term};
  \node[ob=4.6cm] (cl) at (0,-2.1) {a causal link $(a,p)$ where activity $a$ is causally upstream of process $p$};
  \node[ob=2.4cm] (proc) at (0,-4.2) {a process};
  \node[ob=2.0cm] (bp) at (5.0,-4.2) {a BP term};
  \draw[asp] (act) -- node[lab, above] {is enabled by} (gp);
  \draw[asp] (act) -- node[lab, above] {has as type} (mf);
  \draw[asp] (cl) -- node[lab, right] {$a$} (act);
  \draw[asp] (cl) -- node[lab, right] {$p$} (proc);
  \draw[asp] (proc) -- node[lab, above] {has as type} (bp);
\end{tikzpicture}
\end{equation}
together with analogous spans for ``occurs in'' (\lt{an occurrence}, with
aspects to \lt{an activity or a process}, a coproduct, and to \lt{a
location}) and ``is part of'' (\lt{a parthood}, with two aspects to
\lt{a location}). The aspect ``is enabled by'' is functional because, in GO-CAM,
each activity node is enabled by a single gene product or complex; ``causally
upstream of'' is many-to-many and is therefore a span. The model of
Figures~\ref{fig:foxp1-cam} and \ref{fig:foxp1-noctua} is the instance $I$
shown in Table~\ref{tab:gocam}, in which each type becomes a table and each
aspect a column \citep[Section~3]{SpivakKent2012}.

\begin{table}[ht]
\centering\small
\begin{tabular}[t]{@{}lll@{}}
\toprule
\lt{an activity} & is enabled by & has as type \\
\midrule
$a_1$ & Foxp1 & DNA-binding transcription factor activity \\
\bottomrule
\end{tabular}
\hspace{1.2em}
\begin{tabular}[t]{@{}lll@{}}
\toprule
\lt{a causal link} & $a$ & $p$ \\
\midrule
$c_1$ & $a_1$ & $p_1$ \\
$c_2$ & $a_1$ & $p_2$ \\
\bottomrule
\end{tabular}

\vspace{1.2ex}
\begin{tabular}[t]{@{}ll@{}}
\toprule
\lt{a process} & has as type \\
\midrule
$p_1$ & regulation of action potential \\
$p_2$ & innate vocalization behavior \\
\bottomrule
\end{tabular}
\hspace{1.2em}
\begin{tabular}[t]{@{}ll@{}}
\toprule
\lt{a location} & has as type \\
\midrule
$\ell_1$ & corpus striatum (Uberon) \\
$\ell_2$ & medium spiny neuron (Cell Ontology) \\
\bottomrule
\end{tabular}

\vspace{1.2ex}
\begin{tabular}[t]{@{}lll@{}}
\toprule
\lt{an occurrence} & what & where \\
\midrule
$o_1$ & $a_1$ & $\ell_1$ \\
$o_2$ & $p_1$ & $\ell_2$ \\
\bottomrule
\end{tabular}
\hspace{1.2em}
\begin{tabular}[t]{@{}lll@{}}
\toprule
\lt{a parthood} & part & whole \\
\midrule
$h_1$ & $\ell_2$ & $\ell_1$ \\
\bottomrule
\end{tabular}
\caption{The GO-CAM model of Figures~\ref{fig:foxp1-cam} and
\ref{fig:foxp1-noctua} as an instance of the schema olog~(\ref{olog:gocam}).
Each table is a type and each column an aspect; \lt{an occurrence} and
\lt{a parthood} are the spans for ``occurs in'' and ``is part of''.}
\label{tab:gocam}
\end{table}

Standard annotations can be read off such an instance by
composing aspects. For example, each activity $a$ yields the annotation
$\bigl(I(\text{is enabled by})(a),\ \text{enables},\ I(\text{has as type})(a)\bigr)$,
which for $a_1$ is the first row of Table~\ref{tab:annotation}. The
locations, by contrast, are attached to individual activities and processes,
not to Foxp1 itself. This is why corpus striatum and medium spiny neuron do
not appear as GO annotations of Foxp1 in Figure~\ref{fig:foxp1-std}.

\subsection{Comparing Gene ologs}\label{sec:functors}

As with ologs in general, two Gene ologs can be related by a \emph{functor}, a
structure-preserving map from one category to another
\citep[Section~4]{SpivakKent2012}. A functor $F\colon\mathcal{C}\to\mathcal{D}$
sends each type $A$ of $\mathcal{C}$ to a type $F(A)$ of $\mathcal{D}$ and each
aspect $f\colon A\to B$ to an aspect, or more generally a path,
$F(f)\colon F(A)\to F(B)$, in such a way that facts are preserved: equivalent
paths in $\mathcal{C}$ are sent to equivalent paths in $\mathcal{D}$. In
particular $F(g\circ f)=F(g)\circ F(f)$ and identities are preserved
(Figure~\ref{fig:functor}).

\begin{figure}[ht]
\centering
\begin{tikzpicture}
  \node[ob=1.7cm] (A) at (0,0) {type $A$};
  \node[ob=1.7cm] (B) at (3,0) {type $B$};
  \node[ob=1.7cm] (C) at (3,-2) {type $C$};
  \draw[asp] (A) -- node[lab, above] {$f$} (B);
  \draw[asp] (B) -- node[lab, right] {$g$} (C);
  \draw[asp] (A) -- node[lab, below left] {$g\circ f$} (C);
  \draw[-{Stealth[length=6pt]}, thick] (4.6,-1) -- node[lab, above] {$F$} (6.0,-1);
  \node[ob=1.9cm] (FA) at (7.4,0) {type $F(A)$};
  \node[ob=1.9cm] (FB) at (10.6,0) {type $F(B)$};
  \node[ob=1.9cm] (FC) at (10.6,-2) {type $F(C)$};
  \draw[asp] (FA) -- node[lab, above] {$F(f)$} (FB);
  \draw[asp] (FB) -- node[lab, right] {$F(g)$} (FC);
  \draw[asp] (FA) -- node[lab, below left, align=center, pos=0.6] {$F(g\circ f)$\\$=F(g)\circ F(f)$} (FC);
\end{tikzpicture}
\caption{A functor $F$ between two Gene ologs sends types to types and aspects
to aspects (or paths), and preserves composition and facts.}
\label{fig:functor}
\end{figure}

Functors make the comparison between two Gene ologs precise.
Figure~\ref{fig:tor} shows a functor between two small Gene ologs, sending
\lt{a TOR signaling} to \lt{a cAMP-mediated signaling} and acting as the
identity on the other types. It is a valid functor, since it preserves all
the structure, and it shows that the two terms occupy the \emph{same
structural position}: both are intracellular signal transductions. It does
not, however, show that the two terms are equivalent, and it is not
\emph{meaningful} in the sense of \citet[Section~4.3.1]{SpivakKent2012}: a
functor is meaningful when every intended instance of a type $X$ would also
be accepted as an instance of $F(X)$, and a TOR signaling event is not a
cAMP-mediated signaling event. Functors of this kind therefore express
structural analogy. Meaningful functors include the inclusion of a smaller
Gene olog into a larger one, such as the inclusions from a common ground
described next.

\begin{figure}[ht]
\centering
\begin{tikzpicture}
  \node[ob=3.1cm] (t) at (0,0) {a TOR signaling};
  \node[ob=3.1cm] (i1) at (0,-1.5) {an intracellular signal transduction};
  \node[ob=3.1cm] (s1) at (0,-3.0) {a signal transduction};
  \draw[isa] (t) -- node[lab, right] {is an} (i1);
  \draw[isa] (i1) -- node[lab, right] {is a} (s1);
  \draw[-{Stealth[length=6pt]}, thick] (2.4,-1.5) -- node[lab, above] {$F$} (4.0,-1.5);
  \node[ob=3.1cm] (c) at (6.4,0) {a cAMP-mediated signaling};
  \node[ob=3.1cm] (i2) at (6.4,-1.5) {an intracellular signal transduction};
  \node[ob=3.1cm] (s2) at (6.4,-3.0) {a signal transduction};
  \draw[isa] (c) -- node[lab, right] {is an} (i2);
  \draw[isa] (i2) -- node[lab, right] {is a} (s2);
\end{tikzpicture}
\caption{A functor between two Gene ologs that sends \lt{a TOR signaling} to
\lt{a cAMP-mediated signaling} and is the identity on the other types. It
preserves structure but does not preserve meaning.}
\label{fig:tor}
\end{figure}

\subsection{Merging Gene ologs}\label{sec:merge}

When two Gene ologs share a common part, they can be glued together along it.
Formally, a \emph{common ground} for Gene ologs $\mathcal{C}$ and $\mathcal{D}$
is a third Gene olog $\mathcal{B}$ with meaningful functors
$F_{\mathcal{C}}\colon\mathcal{B}\to\mathcal{C}$ and
$F_{\mathcal{D}}\colon\mathcal{B}\to\mathcal{D}$
\citep[Section~4.3.1]{SpivakKent2012}. The merge of $\mathcal{C}$ and
$\mathcal{D}$ along $\mathcal{B}$ is the pushout
$\mathcal{C}\sqcup_{\mathcal{B}}\mathcal{D}$ of the span
$\mathcal{C}\xleftarrow{F_{\mathcal{C}}}\mathcal{B}\xrightarrow{F_{\mathcal{D}}}\mathcal{D}$,
taken in the category of ologs (categories): it contains everything in
$\mathcal{C}$ and $\mathcal{D}$, with the images of $\mathcal{B}$ identified.
For the two Gene ologs of Figure~\ref{fig:tor}, the common ground $\mathcal{B}$
is the olog
$\lt{an intracellular signal transduction}\hookrightarrow\lt{a signal transduction}$,
included in both, and the pushout is shown in Figure~\ref{fig:merge}. Note that
the merge is built from the inclusions of $\mathcal{B}$, not from the functor
$F$ of Figure~\ref{fig:tor}; using $F$ would identify TOR signaling with
cAMP-mediated signaling.

\begin{figure}[ht]
\centering
\begin{tikzpicture}
  \node[ob=3.1cm] (t) at (-2.9,0) {a TOR signaling};
  \node[ob=3.1cm] (c) at (2.9,0) {a cAMP-mediated signaling};
  \node[ob=3.1cm] (i) at (0,-2.0) {an intracellular signal transduction};
  \node[ob=3.1cm] (s) at (0,-3.6) {a signal transduction};
  \draw[isa] (t.south) -- node[lab, below left] {is an} (i.north west);
  \draw[isa] (c.south) -- node[lab, below right] {is an} (i.north east);
  \draw[isa] (i) -- node[lab, right] {is a} (s);
  \draw[dashed, rounded corners] (-2.0,-1.3) rectangle (2.0,-4.15);
  \node[font=\footnotesize, anchor=west] at (2.05,-3.95) {image of the common ground $\mathcal{B}$};
\end{tikzpicture}
\caption{The merge of the two Gene ologs of Figure~\ref{fig:tor} along their
common ground $\mathcal{B}$ (dashed box), computed as a pushout of ologs.}
\label{fig:merge}
\end{figure}

Requiring the common part to be literally identical is a special case. In
general, $\mathcal{B}$ only needs to map meaningfully into $\mathcal{C}$ and
$\mathcal{D}$, so two Gene ologs (for example, ologs built by different
curation groups or for different species) can communicate through a shared
common ground, and the common ground can be expanded over time
\citep{SpivakKent2012}. Instance data can also be transferred along functors
\citep{Spivak2012}.

\section{Discussion: Gene ologs and OWL}\label{sec:discussion}

The examples above suggest what an olog view adds to the standard
representation of GO:
\begin{itemize}
  \item \emph{Functionality is made explicit.} Every aspect must be a
  function. Writing a Gene olog therefore forces the modeller to decide, for
  each GO relation, whether it is functional (and can be composed) or
  many-to-many (and must be represented as a span). GO's ``all--some''
  relations do not make this distinction.
  \item \emph{Path equations are first-class.} Facts such as
  olog~(\ref{olog:dogma}) and the annotation constraint in
  olog~(\ref{olog:annotation}) are part of the ontology, and they become
  integrity constraints on instance data. OWL~2 property chains can express
  some related constraints, but equality of paths is not a primitive notion
  in OWL.
  \item \emph{Constructed types have fixed meaning.} Pullbacks, pushouts,
  products and coproducts define new types by universal properties, as in
  olog~(\ref{olog:pb-go}) and olog~(\ref{olog:po-go}).
  \item \emph{Data and comparison are built in.} Instance data are functors to
  $\mathbf{Set}$, so annotations and GO-CAM models live in the same framework
  as the ontology, and functors and pushouts give a precise account of
  comparing and merging ontologies.
\end{itemize}

There are also clear limitations. Ologs, as used here, are two-valued, like
OWL, so they do not address uncertain or graded knowledge. Aspects must be
total functions, so optional or partial information (for example, an activity
whose location is unknown) must be modelled with spans or with suitably
restricted types, which can make Gene ologs verbose. The negation (``NOT'')
qualifier of GO annotations, and OWL constructs such as universal
restrictions and disjointness axioms, have not been addressed here. Finally,
we have not studied the computational cost of reasoning with Gene ologs at
the scale of GO, so no claim about scalability relative to OWL is made. For
these reasons we view Gene ologs as complementary to the OWL representation of
GO: a way to make the compositional structure of GO knowledge explicit, rather
than a replacement for the existing OWL infrastructure and reasoners.

\section{Conclusion and future work}\label{sec:conclusion}

Category theory has been advocated as a language for forming precise
connections between ideas across the sciences \citep{Spivak2014}. Ologs, which
are rooted in category theory, provide a framework for codifying knowledge
that is both flexible and rigorous. In this work we have taken an initial step
towards applying ologs to GO, one of the largest stores of biological
knowledge. We introduced Gene ologs, i.e.\ ologs for GO; fixed the reading of
GO terms as types whose instances are occurrences of processes and activities
and particular cellular structures; examined which GO relations qualify as
aspects; illustrated types, aspects, facts, instances, products, coproducts,
pullbacks and pushouts with GO examples; represented standard annotations and
GO-CAM models as instance data subject to checkable facts; and described how
Gene ologs can be compared by functors and merged by pushouts.

Future work includes a systematic translation between the OWL axioms of GO and
Gene ologs; the treatment of negated annotations and other OWL constructs;
larger GO-CAM case studies, including the use of functorial data migration to
derive standard annotations from GO-CAM models; and an implementation that
checks the facts of a Gene olog against GO release data.

\section*{Acknowledgements}
The author thanks Claude Opus 5.5 (Anthropic) for assistance in reviewing the
first version of this manuscript, identifying logical gaps and preparing this
revision.

\end{document}